\documentclass[lettersize,journal]{IEEEtran}
\usepackage{amsmath,amsfonts}
\usepackage{algorithmic}
\usepackage{algorithm}
\usepackage{array}
\usepackage[caption=false,font=normalsize,labelfont=sf,textfont=sf]{subfig}
\usepackage{textcomp}
\usepackage{stfloats}
\usepackage{url}
\usepackage{verbatim}
\usepackage{graphicx}
\usepackage{cite}
\usepackage{xspace}
\usepackage{amsmath}
\begin{document}

\newcommand{\sysname}{HURRY\xspace}
\title{\sysname: \underline{H}ighly \underline{U}tilized, \underline{R}econfigurable \underline{R}eRAM-based In-situ Accelerator with Multifunctionalit\underline{y}}

\author{Hery Shin,~\IEEEmembership{Member,~IEEE,} Jae-Young Kim,~\IEEEmembership{Member,~IEEE,} Donghyuk Kim,~\IEEEmembership{Member,~IEEE,} Joo-Young Kim,~\IEEEmembership{Senior Member,~IEEE}
\thanks{Manuscript received June 28, 2024.}}

\markboth{Journal of \LaTeX\ Class Files,~Vol.~14, No.~8, August~2024}%
{Shell \MakeLowercase{\textit{et al.}}: A Sample Article Using IEEEtran.cls for IEEE Journals}

\IEEEpubid{0000--0000/00\$00.00~\copyright~2024 IEEE}

\maketitle

\begin{abstract}

Resistive random-access memory (ReRAM) crossbar arrays are suitable for efficient inference computations in neural networks due to their analog general matrix-matrix multiplication (GEMM) capabilities. However, traditional ReRAM-based accelerators suffer from spatial and temporal underutilization. We present \sysname, a reconfigurable and multifunctional ReRAM-based in-situ accelerator. \sysname uses a block activation scheme for concurrent activation of dynamically sized ReRAM portions, enhancing spatial utilization. Additionally, it incorporates functional blocks for convolution, ReLU, max pooling, and softmax computations to improve temporal utilization. System-level scheduling and data mapping strategies further optimize performance. Consequently, \sysname achieves up to 3.35$\times$ speedup, 5.72$\times$ higher energy efficiency, and 7.91$\times$ greater area efficiency compared to current ReRAM-based accelerators.

\end{abstract}


\section{Introduction}

\IEEEPARstart{M}{achine} learning algorithms, particularly deep and convolutional neural networks (DNNs and CNNs), are essential for applications ranging from mobile devices to cloud data centers. These algorithms require substantial storage for weights and frequent weight access for tasks such as convolution, necessitating specialized hardware like neural processing units and in-memory processors. Among these, ReRAM-based in-situ accelerators (RIAs) for CNN inference stand out for their cell density and energy-efficient GEMM processing capabilities.~\cite{yu2021rram, chi2016prime, shafiee2016isaac}.

However, existing RIAs exhibit low utilization of ReRAM arrays, which diminishes their storage and computational efficiency. We examine this issue from two perspectives: spatial and temporal utilization. Spatial utilization is the average ratio of data-mapped cells to the ReRAM array per CNN layer, while temporal utilization is the average ratio of activated ReRAM cells to the ReRAM array per clock cycle.

\begin{figure}[t]
  \centering
    \includegraphics[width=0.47\textwidth]
    {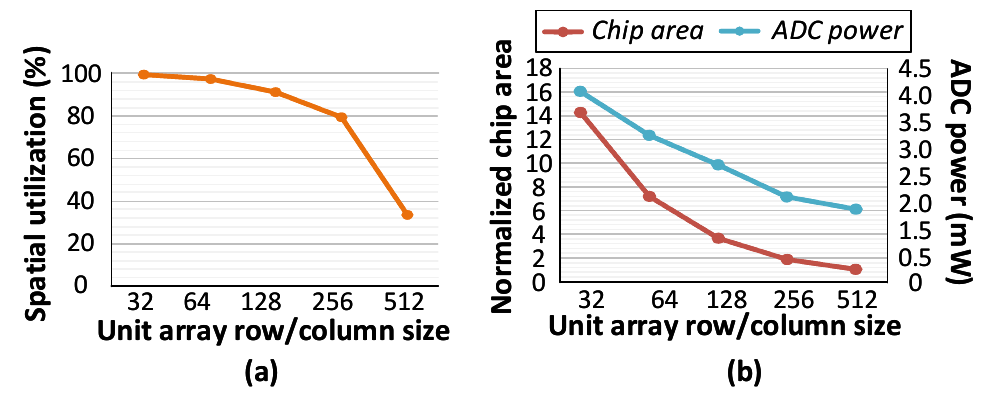}
  \vspace{-0.1in}
  \caption{(a) Unit array size vs. ReRAM spatial utilization rate (b) Unit array size vs. Chip size/ADC power consumption}
  \vspace{-0.2in}
  \label{figure1}
\end{figure}

Firstly, the trade-off between ReRAM spatial utilization and peripheral overhead, such as analog-to-digital converters (ADCs) and digital-to-analog converters (DACs), results in suboptimal ReRAM array sizes, compromising energy and area efficiency. Experiments with ISAAC~\cite{shafiee2016isaac} using PUMAsim~\cite{ankit2019puma} on AlexNet illustrate this impact. As shown in Fig.~\ref{figure1}(a), spatial utilization drops from 99\% to 57\% as array size increases from 128$\times$128 to 512$\times$512, indicating that larger arrays than CNN kernels lead to underutilization. Thus, previous studies~\cite{chi2016prime, shafiee2016isaac, 9866551} have opted for smaller arrays like 128$\times$128 to maintain high spatial utilization.

However, smaller arrays decrease energy and area efficiency due to higher peripheral overhead. ADCs, in particular, contribute over 60\% of power and area consumption in RIAs~\cite{chi2016prime, shafiee2016isaac}. Fig.~\ref{figure1}(b) shows that 16 units of 128$\times$128 arrays with 7-bit ADCs consume 3.4$\times$ more ADC power and have a 3.7$\times$ higher area overhead compared to a 512$\times$512 array with a 9-bit ADC. Thus, designing a system that maximizes spatial utilization while minimizing peripheral overhead is crucial for high performance.

Secondly, the frequent data movement between ReRAM arrays and computing units drastically reduces the temporal utilization, constituting up to 48\% of runtime in ISAAC~\cite{shafiee2016isaac}. Existing RIAs focus primarily on GEMM for convolutional layers ($Conv$) ~\cite{chi2016prime, shafiee2016isaac, zhu2018mixed}, neglecting other CNN operations like max pooling ($Max$), ReLU ($ReLU$), and residual ($Res$) layers. This results in extensive data movement between operators and idle ReRAM arrays, decreasing temporal utilization.

\IEEEpubidadjcol

In short, previous designs trade high spatial utilization for high peripheral overhead and exhibit reduced performance due to data movement, impacting temporal utilization. To address these issues, we introduce \sysname, a reconfigurable, multifunctional ReRAM-based in-situ accelerator. \sysname uses large reconfigurable ReRAM arrays to balance spatial utilization and peripheral overhead, while its multifunctional arrays reduce data movement, enhancing temporal utilization. Unlike previous works such as MISCA ~\cite{zhu2018mixed}, which focus on reconfigurability for spatial utilization, to the best of our knowledge, our work is the first to integrate both reconfigurability and multifunctionality into ReRAM arrays, aiming to improve both spatial and temporal utilizations. The major contributions of this paper are as follows:

\begin{itemize}
\item We identify spatial and temporal underutilization issues in RIAs and propose \sysname, featuring reconfigurability and multifunctionality.
\item We develop a reconfigurable and multifunctional ReRAM that dynamically resizes and activates portions at runtime to increment spatial utilization and computes various CNN layers to enhance temporal utilization.
\item We establish system-level scheduling and functional block-level data mapping to improve overall ReRAM utilization.
\item Evaluations with various CNN benchmarks show that \sysname achieves up to 3.35$\times$ speedup, 5.72$\times$ greater energy efficiency, and 7.91$\times$ higher area efficiency compared to the baselines.
\end{itemize}
\vspace{0.1in}
\section{\sysname Architecture}

\subsection{Overview}
\begin{figure}[t]
\centering
\includegraphics[width=0.48\textwidth]{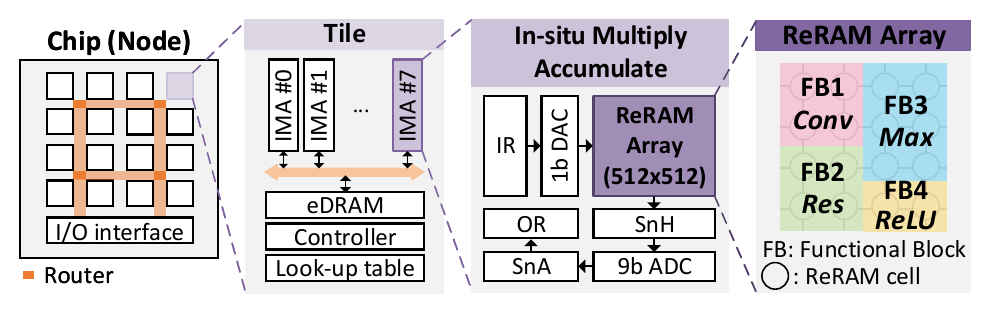}
\vspace{-0.1in}
\caption{Overall architecture of \sysname}
\vspace{-0.1in}
\label{figure3}
\end{figure}

Fig.~\ref{figure3} illustrates the overall architecture of \sysname. Similar to our baseline ISAAC~\cite{shafiee2016isaac}, a chip includes 16 tiles, each containing 8 in-situ multiply accumulate (IMAs), a 512KB embedded DRAM (eDRAM), a controller, and a look-up table. Each IMA has a 512$\times$512 ReRAM array, 32KB/2KB SRAM for input/output registers (IR/OR), 1-bit DACs, 9-bit ADCs, sample-and-hold units (SnH), and shift-and-add units (SnA). We chose 512$\times$512 as the unit array size of ReRAM, the maximum size with negligible leakage current demonstrated by~\cite{hu2016dot} and in Section~\ref{section_5}. A ReRAM array incorporates functional blocks (FBs) to implement reconfigurability and multifunctionality. Reconfigurability (Section~\ref{subsection3-2}) combines the benefits of small and large arrays, improving spatial utilization and reducing peripheral overhead. Multifunctionality (Section~\ref{subsection3-3}) reduces data movement between ReRAM and computing units, boosting temporal utilization and allowing the omission of output registers and digital computing units within tiles.

\subsection{Block Activation Scheme}
\label{subsection3-2} 

We propose the block activation scheme (BAS) for reconfigurability, adapting the third-voltage selecting scheme~\cite{zackriya2017novel}. While $V_{set}$ is used for writing, 1/3$V_{set}$ or 2/3$V_{set}$ is applied to non-written cells to suppress sneak-path currents. BAS uses these voltages for reading, enabling concurrent operations and dynamic FB size adjustments through reconfigurable WL and BL configurations.

\begin{figure}[t]
  \centering
  \includegraphics[width=0.48\textwidth]{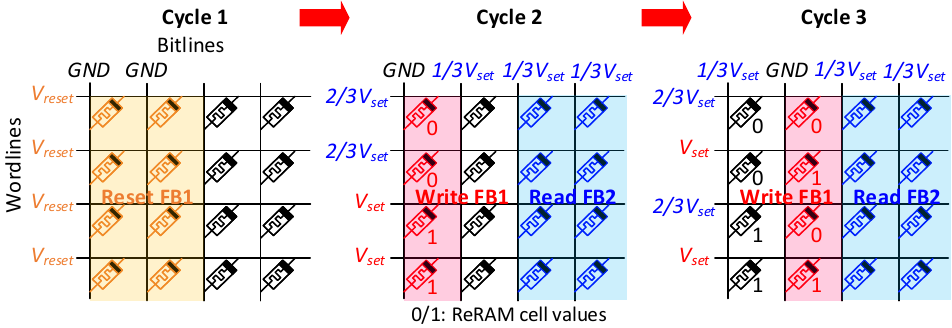}
  \vspace{-0.05in}
  \caption{Block activation scheme simultaneously writing FB1 and reading FB2}
  \vspace{-0.1in}
  \label{figure4}
\end{figure}

Fig.~\ref{figure4} illustrates BAS in a 4$\times$4 ReRAM array with two 4$\times$2 FBs. In cycle 1, FB1 is reset for writing by applying $V_{reset}$ to WL and grounding BL (GND). In cycle 2, FB1's first column is written with '1' using $V_{set}$/GND and '0' with 2/3$V_{set}$/GND. Columns not designated for writing and FBs for reading set their BLs to 1/3$V_{set}$. FB2 continues reading with 1/3$V_{set}$ or 2/3$V_{set}$ applied. The process repeats for the second column in cycle 3. Writing and reading require cycles equal to the columns in the FB.

ReRAM cell precision is set to one bit for three reasons: 
1) the third-voltage scheme assumes $V_{read}$ is one-third or two-thirds of $V_{set}$, incompatible with multi-level cells where lower bit $V_{set}$ might fall below higher bit $V_{read}$,
2) physically implemented ReRAM chips are limited to one or two-bit precision due to process variations and read noise, which is more pronounced in larger arrays ~\cite{9062953}. One-bit cells are more resilient to these nonidealities, and
3) it simplifies BAS implementation, requiring control over only four voltage levels ($V_{set}$, $1/3V_{set}$, $2/3V_{set}$, and GND).
Reduced computational density due to lower cell precision is mitigated by using larger unit arrays like 512$\times$512 with BAS, as shown in Section~\ref{section_5}.

\subsection{Functional Block Implementation}
\label{subsection3-3}

\sysname exploits multifunctionality by enabling FBs to execute various CNN layer functions using on-chip ReRAM's storage, VMM, and in-memory logic capabilities. Weights are mapped onto the ReRAM array for $Conv$ layers and inputs for other layers, as detailed in Section~\ref{subsection4-3}.

\subsubsection{Conv, Res, and FC}
The $Conv$ FB primarily executes GEMM operations. The $Res$ FB handles residual layers by adding the $Conv$ FB's GEMM output to its input, facilitated by accumulating current along the BLs. This allows merging the $Conv$ and $Res$ FBs, as shown in Fig.~\ref{figure5}(a), where the $Res$ FB is positioned underneath the $Conv$ FB for simultaneous reading of GEMM outputs and $Conv$ layer inputs. The fully connected ($FC$) FB also executes GEMM within ReRAM, similar to the $Conv$ FB.

\subsubsection{Max and ReLU}
\begin{figure}[t]
\centering
\includegraphics[width=0.47\textwidth] {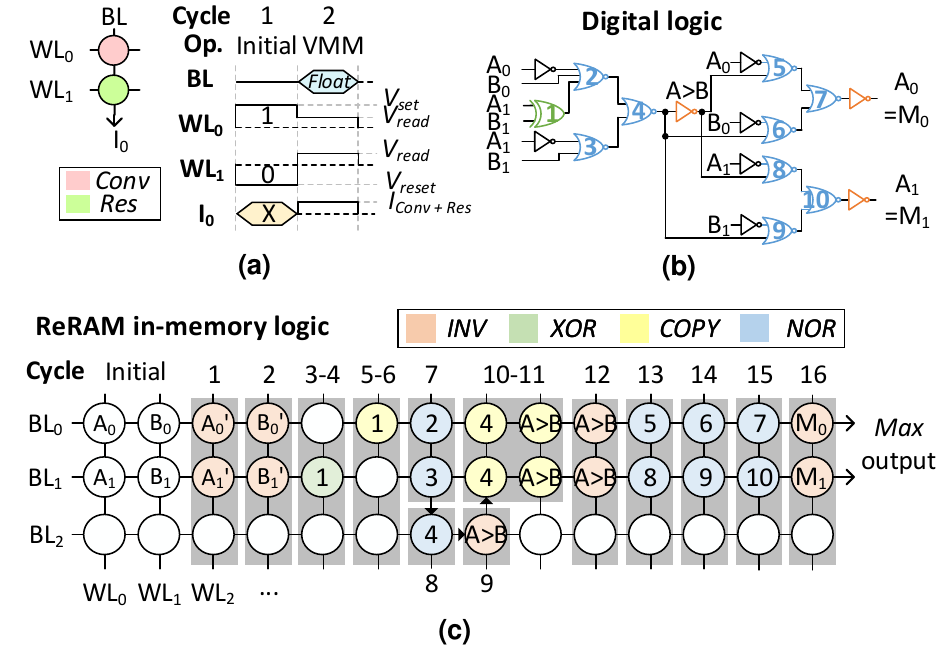}
\vspace{-0.05in}
\caption{Functional block implementation (a) Combined convolutional and residual functional block and (b) max pooling functional block}
\vspace{-0.1in}
\label{figure5}
\end{figure}

The $Max$ FB executes a step-wise tournament of compare and select operations, known as "max logic," as illustrated in Fig.~\ref{figure5}(b). We refer to~\cite{yang2020retransformer} and~\cite{kvatinsky2014magic} for the max logic implementation. Fig.~\ref{figure5}(c) shows an example of max logic in ReRAM for comparing two 2-bit elements, A and B, involving 11 cycles for comparison and 5 cycles for select logic. The $ReLU$ function uses max logic, including zero in the comparison. The $ReLU$ FB can combine with the $Max$ FB to reduce runtime and control logic complexity. Efficient FB merging strategies are discussed in Section~\ref{section_4}.

\subsubsection{Softmax Support}
The $softmax$ FB uses max logic, following the scheme proposed by~\cite{yang2020retransformer}. According to equation (1), after identifying the maximum element \( x_{\text{max}} \), the $softmax$ calculation reduces to a single exponential and logarithmic operation, offloaded to look-up table circuits within the tile.

\begin{align}
    \begin{split}
        \textit{Softmax}(x_i) &= \frac{e^{x_i}}{\sum_{j=1}^{n} e^{x_j}} = \frac{e^{x_i - x_{\text{max}}}}{\sum_{j=1}^{n} e^{x_j - x_{\text{max}}}} \\
        &= \exp\left( x_i - x_{\text{max}} - \log\left(\sum_{j=1}^{n} e^{x_j - x_{\text{max}}}\right) \right)
    \end{split}
\end{align}

This approach reduces power consumption and minimizes data movement compared to arithmetic logic units (ALUs), as shown in Section~\ref{section_5}.

\section{Model-aware Scheduling}
\label{section_4}

\subsection{Inter-FB Scheduling}
At the chip level, \sysname receives inputs through an I/O interface to individual IMAs, each configured for different layers. FBs within an IMA process results, which are aggregated in the OR after SnA operations through ADCs. Outputs are released only after an IMA completes all layer computations, with softmax completed via a lookup table. This flow reduces data movement and enhances temporal utilization.

\begin{figure}[t]
  \centering
  \includegraphics[width=0.46\textwidth]{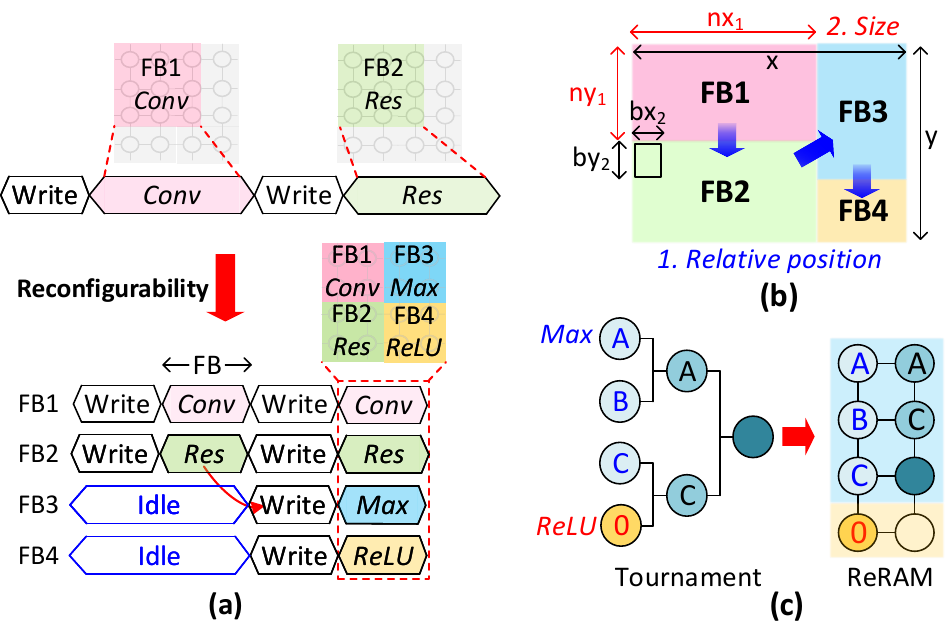}
  \vspace{-0.1in}
  \caption{(a) Fine-grained pipelining of FBs (b) Relative positioning and sizing of FBs within a ReRAM array (c) Max pooling and ReLU FBs merged by data mapping strategy}
  \vspace{-0.1in}
  \label{figure6}
\end{figure}

Within each IMA, a fine-grained pipeline operates at the FB level, activating multiple FBs simultaneously to increase spatial utilization, as shown in Fig.~\ref{figure6}(a). For example, while an output of combined $Conv$ and $Res$ FB is being written to a $Max$ FB, the next $Conv$ FB inputs can be loaded simultaneously. This synchronization minimizes idle times, which occur before the initial $Conv$ FB output is ready for the $Max$ FB. In the first layer of AlexNet, the $Conv$ FB operates for 196 cycles, while the combined $Max$ and $ReLU$ FBs take 168 cycles within the IMA, demonstrating the efficiency of tightly pipelined FBs with minimal idle time.

\subsection{Inter-FB Mapping}
\label{section_4_2}

Building on insights from Section~\ref{subsection3-3}, strategic FB placement and sizing are crucial to reduce computation latency and simplify control complexity.

\vspace{-0.1in}

\begin{algorithm}[H]
\small
\caption{FB Relative Positioning Algorithm}
\label{alg:alg1}
\begin{algorithmic}
\STATE 
\STATE \textbf{Input:} The CNN model
\STATE \textbf{Output:} Sequence pair of FBs
\STATE 
\STATE $n$: the number of kernels in convolutional layers
\STATE $i$: an identifier of $i$th FB
\STATE $seq1$: first sequence of identifiers
\STATE $seq2$: second sequence of identifiers
\STATE \textbf{Initialize:} $i = 1$, $seq1 = (1)$, $seq2 = (1)$
\STATE 
\WHILE{$i < n+1$}
    \STATE $j = 1$ \COMMENT{Initialize the inner loop counter}
    \WHILE{$j < i$}
        \IF{$i$th FB involves accumulative operations with $j$th FB}
            \STATE Place $i$ left to $j$ in the $seq2$ 
        \ELSE
            \STATE $k = $ the rightmost number in $seq1$
            \STATE Place $i$ on the far right in $seq1$
            \STATE Place $i$ left to $k$ in the $seq2$ 
        \ENDIF
        \STATE Increment $j$ by 1
    \ENDWHILE
    \STATE Increment $i$ by 1
\ENDWHILE
\end{algorithmic}
\label{alg1}
\end{algorithm}


\subsubsection{FB Relative Positioning}
The first step is identifying which FBs to merge and their relative positioning, as shown in Fig.~\ref{figure6}(b)-1. Algorithm 1 handles FB positioning, arranging them horizontally (left/right) or vertically (above/below) using a sequence pair representation~\cite{552084}. For example, if FB2 uses FB1's output, it is placed below FB1, with FB2's identifier to the left of FB1's in the second sequence. Otherwise, FB2 is placed to the right of FB1, with its identifier after FB1's in the first sequence.

\subsubsection{FB Size Balancing} 
The next step involves sizing each FB, illustrated in Fig.~\ref{figure6}(b)-2, to balance workloads, avoid stalls, and eventually enhance temporal utilization. Algorithm 2 uses a greedy method to determine the optimal FB size, ensuring that each FB's computational output does not exceed that of its predecessors and that all FBs collectively fit within the total array size.

\vspace{-0.1in}
\begin{algorithm}[H]
\small
\caption{FB Size Balancing Algorithm}
\label{alg:alg2}
\begin{algorithmic}
\STATE 
\STATE \textbf{Input:} The CNN model, unit ReRAM array size, a list of the required sizes of operations
\STATE \textbf{Output:} Row and column sizes of FBs
\STATE 
\STATE $(nx_i, ny_i)$: the size of the $i$th FB
\STATE $(bx_i, by_i)$: the required size of an operation in the $i$th FB
\STATE $(x, y)$: the row and column size of a unit ReRAM array
\STATE \textbf{Initialize:} $nx_i = x$, $ny_i = y$
\STATE 
\FOR{$i = 2$ to $nx$}
    \STATE
    \[
    \begin{split}
    nx_i = \underset{nx_i}{\arg\max} \Big\{ & \left( \sum_i nx_i \leq arr_x \right) \cap \left( \sum_i ny_i \leq arr_y \right) \\
    & \cap \left( \frac{nx_{i-1}}{bx_{i-1}} \times \frac{ny_{i-1}}{by_{i-1}} \leq \frac{ny_i}{by_{i-1}} \right) \Big\}
    \end{split}
    \]
    \STATE Update $(nx_i, ny_i)$
    \STATE Increment $i$ by 1
\ENDFOR
\end{algorithmic}
\label{alg2}
\end{algorithm}
\vspace{-0.1in}

\subsection{Intra-FB Data Mapping}
\label{subsection4-3}

To facilitate multifunctionality, we adopt a hybrid mapping scheme (HMS), combining weight and input stationary dataflows. The $Conv$ FB uses a weight stationary dataflow, while other FBs like $Res$, $Max$, and $ReLU$ use an input stationary dataflow.

In the $Conv$ FB, inputs are flattened into column vectors and multiplied by the kernel loaded onto ReRAM arrays. For the $Max$ FB, the $Conv$ FB result is written onto the ReRAM and organized into a tree-shaped tournament in a rectangular array, as shown in Fig.~\ref{figure6}(c). This leverages the principle that the final tree layer contains more leaves than previous layers combined. Consequently, each FB's column count aligns with the leaf count in the tree's final layer.

HMS in a ReRAM array serves three purposes:
1) it mitigates low spatial utilization in weight stationary arrays due to kernel-array column size mismatches,
2) it efficiently implements multifunctionality in FBs that require only layer inputs, not kernels, and
3) it balances kernel reuse and high utilization rates, reducing data movement for kernel retrieval and alleviating concerns about device reliability degradation from frequent writing.
The overhead of the complex controller required to implement HMS is marginal, as further discussed in Section~\ref{section_5}.

\section{Evaluation \& Discussion}
\label{section_5}

\subsection{Experimental Setup}
\subsubsection{Hardware Simulation} 
We modify the PUMAsim open-source simulator~\cite{ankit2019puma} for \sysname's evaluation. Simulator parameters are based on Verilog and synthesized using Synopsys Design Compiler with a TSMC 40 nm standard cell library scaled to a 32 nm process for comparison. The operating frequency is set to 100 MHz. The ReRAM cell energy and area model is based on~\cite{hu2016dot}, consistent with our baseline. The feasibility of BAS in a 512$\times$512 ReRAM array is confirmed through SPICE simulations at ${25}^\circ$C, accounting for thermal noise in memristors, shot noise in circuits, and random telegraph noise in the crossbar.


\subsubsection{Benchmark}  
\sysname is evaluated using three CNN models: AlexNet~\cite{krizhevsky2012imagenet}, VGG-16~\cite{simonyan2014very}, and ResNet-18~\cite{he2016deep}, applied to the CIFAR-10 dataset. Inputs and weights for $Conv$ layers are quantized to 8-bit integers, while $softmax$ inputs and weights are quantized to 16-bit floating-point format.

\begin{figure}[t]
\centering
\includegraphics[width=0.48\textwidth] {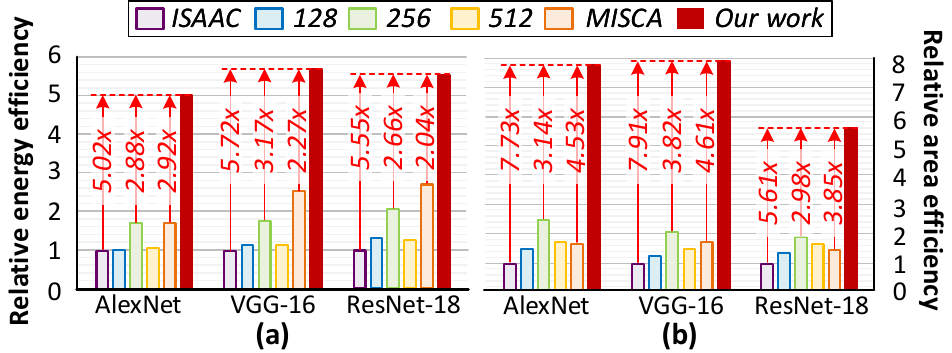}
\vspace{-0.05in}
\caption{(a) Relative energy efficiency with respect to ISAAC and (b) relative area efficiency with respect to ISAAC}
\vspace{-0.1in}
\label{figure7}
\end{figure}

\subsubsection{Baseline} 
We compare \sysname with two baselines: ISAAC~\cite{shafiee2016isaac} and the mixed-size crossbar CNN accelerator MISCA~\cite{zhu2018mixed}. ISAAC is our architectural baseline, and MISCA offers similar reconfigurability by integrating three static ReRAM sizes per IMA. Additionally, we compare with adjusted ISAAC's architectures that match \sysname's configurations but with varying unit array sizes (128$\times$128, 256$\times$256, 512$\times$512), maintaining the same total array size per IMA (16, 4, and 1 array per IMA, respectively). Note that while all compared architectures operate with 2-bit cell precision and perform only GEMM using static-sized ReRAM arrays, \sysname utilizes single-bit precision with multifunctional, reconfigurable-sized ReRAM arrays.

\subsection{Experimental Results}
\label{subsection5-2}

\subsubsection{Efficiency} 
Fig.~\ref{figure7} illustrates the relative energy and area efficiencies of different architectures compared to ISAAC. The 128$\times$128 arrays show low efficiencies due to high ADC overhead, while 512$\times$512 arrays suffer from low spatial and temporal utilization. \sysname significantly outperforms these baselines by 2.66-5.72$\times$ in energy efficiency and 2.98-7.91$\times$ in area efficiency. This improvement is due to enhanced ReRAM utilization. Additionally, \sysname's use of low-power ReRAM for all layers significantly boosts energy efficiency.

\subsubsection{Speedup}
\begin{figure}[t]
\centering
\includegraphics[width=0.48\textwidth] {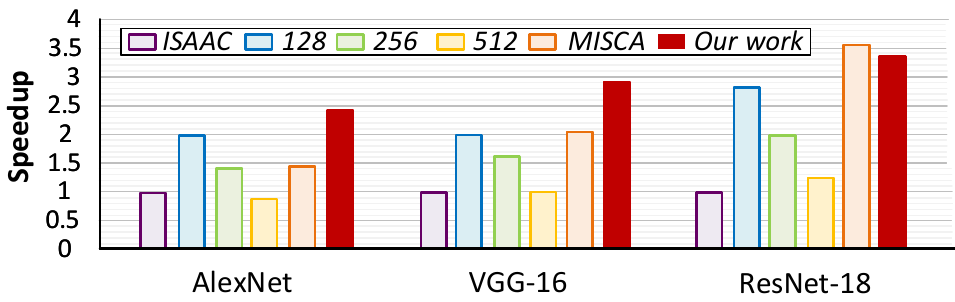}
\caption{Speedup with respect to ISAAC}
\vspace{-0.2in}
\label{figure8}
\end{figure}

Fig.~\ref{figure8} shows \sysname's speed improvement over ISAAC, achieving a 1.21-3.35$\times$ speedup by fully utilizing ReRAM spatially and temporally, with a marginal average accuracy drop of 1.86\% compared to 512$\times$512 arrays. For ResNet-18, \sysname's smaller FB configuration for $Conv$ operations results in a modest speedup compared to MISCA. However, the substantial energy efficiency gains from using multifunctional ReRAM outweigh this speed difference.

\subsubsection{Utilization rate} 
\begin{figure}[t]
\centering
\includegraphics[width=0.48\textwidth] {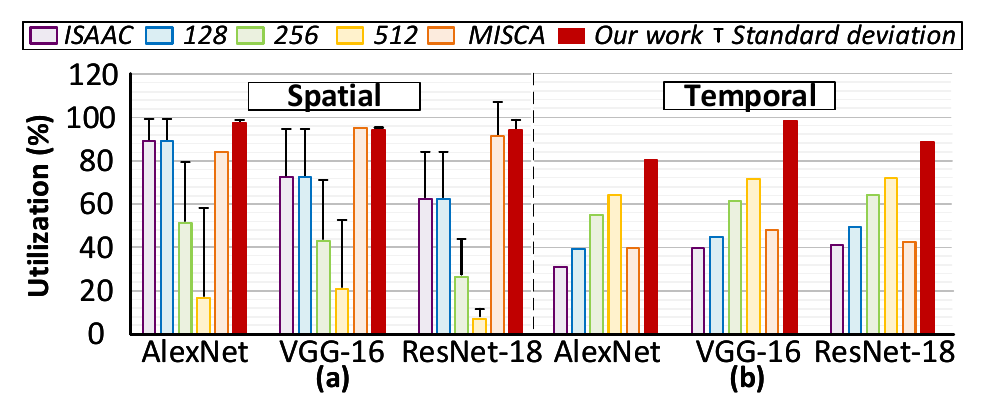}
\vspace{-0.1in}
\caption{(a) Spatial and (b) temporal utilization of ReRAM}
\vspace{-0.1in}
\label{figure9}
\end{figure}

Fig.~\ref{figure9} highlights the spatial and temporal utilization for different CNN models using \sysname. \sysname achieves a 64.17-81.16\% increase in spatial utilization over 512×512 arrays, indicating effective trade-off balancing. \sysname also shows the most consistent spatial utilization across CNN layers with the lowest standard deviation, suggesting its reconfigurable ReRAM adeptly tailors FB configurations to each layer. In contrast, MISCA exhibits greater variations due to its reliance on fixed-size array adjustments.

\sysname improves temporal utilization by 46.72-58.42\% over ISAAC, thanks to its multifunctional ReRAM reducing overall data movement and HMS decreasing inter-IMA data transfers. Remarkably, \sysname also outperforms MISCA by 40.51-50.21\% in temporal utilization. While MISCA prioritizes spatial efficiency for a selected array size, often leaving others idle, \sysname ensures better overall use of resources.

\subsubsection{Overhead} 
The proposed designs require extra storage within an IMA for intermediate results due to ReRAM arrays' larger capacities. \sysname's OR capacity is double that of ISAAC, using 0.0014 $mm^2$ per unit and occupying 1.96\% of the IMA area, with power consumption increased to 0.46 $mW$ or 0.3\% of IMA power. Despite this, a total chip area reduction of 2.6$\times$ offsets the added costs, thanks to reduced ADC overhead and improved energy efficiency.

The controller logic complexity has increased in managing various WL and BL configurations across IMAs. However, compared to static ReRAM setups, the controller's impact on power and area is reasonable for enhanced functionality, accounting for up to 3.35\% of the total power and 12\% of the chip area.

\section{Conclusion}

This paper presents \sysname, a ReRAM-based in-situ accelerator designed to enhance spatial and temporal utilization of conventional RIAs through its reconfigurable and multifunctional ReRAM array architecture. Reconfigurability, implemented via BAS, balances spatial utilization and ADC overhead, while the multifunctional ReRAM array improves temporal utilization by reducing data movement. Combined with advanced system-level scheduling and data mapping strategies, \sysname significantly boosts utilization—spatial by 25.79\% and temporal by 58.42\%. As a result, \sysname achieves up to 3.35$\times$ better performance, 5.72$\times$ higher energy efficiency, and 7.91$\times$ greater area efficiency compared to existing baselines.


\bibliographystyle{IEEEtran}
\bibliography{References.bib}

\vfill

\end{document}